\documentstyle[12pt]{article}
\include{epsf}
 
\newlength{\extralineskip}
\newcommand{\beq}{\begin{equation}}
\newcommand{\eeq}{\end{equation}}
\newcommand{\bd}{\begin{displaymath}}
\newcommand{\ed}{\end{displaymath}}
\def\bea{\begin{eqnarray}}
\def\eea{\end{eqnarray}}

\def\ba{\beq\new\begin{array}{c}}
\def\ea{\end{array}\eeq}

\def\inbar{\,\vrule height1.5ex width.4pt depth0pt}
\def\IC{\relax\hbox{$\inbar\kern-.3em{\rm C}$}}
\def\IR{\relax{\rm I\kern-.18em R}}
\def\IN{\relax{\rm I\kern-.18em N}}

\def\Tr{{\rm Tr}}

\def\e{~{\rm e}}
\makeatletter
\newdimen\normalarrayskip              
\newdimen\minarrayskip                 
\normalarrayskip\baselineskip
\minarrayskip\jot
\newif\ifold             \oldtrue            \def\new{\oldfalse}
\def\arraymode{\ifold\relax\else\displaystyle\fi} 
\def\@arrayskip{\ifold\baselineskip\z@\lineskip\z@
     \else
     \baselineskip\minarrayskip\lineskip2\minarrayskip\fi}
\def\@arrayclassz{\ifcase \@lastchclass \@acolampacol \or
\@ampacol \or \or \or \@addamp \or
   \@acolampacol \or \@firstampfalse \@acol \fi
\edef\@preamble{\@preamble
  \ifcase \@chnum
     \hfil$\relax\arraymode\@sharp$\hfil
     \or $\relax\arraymode\@sharp$\hfil
     \or \hfil$\relax\arraymode\@sharp$\fi}}
\def\@array[#1]#2{\setbox\@arstrutbox=\hbox{\vrule
     height\arraystretch \ht\strutbox
     depth\arraystretch \dp\strutbox
     width\z@}\@mkpream{#2}\edef\@preamble{\halign \noexpand\@halignto
\bgroup \tabskip\z@ \@arstrut \@preamble \tabskip\z@ \cr}%
\let\@startpbox\@@startpbox \let\@endpbox\@@endpbox
  \if #1t\vtop \else \if#1b\vbox \else \vcenter \fi\fi
  \bgroup \let\par\relax
  \let\@sharp##\let\protect\relax
  \@arrayskip\@preamble}

\begin{document}
\thispagestyle{empty}
 
\vskip2cm
\begin{center}
{\huge \bf Order Parameter for Confinement in Large $N$ 
Gauge Theories with Fundamental Matter}\\
\vskip1cm
{\bf   L.D. Paniak}\footnote{Work supported in part by a University of 
British Columbia Graduate Fellowship.}\\
\bigskip

{\it Department of Physics and Astronomy,\\
University of British Columbia\\6224 Agricultural Road\\
Vancouver, British Columbia, Canada V6T 1Z1}
\vskip4cm
\begin{abstract}
In a solvable model of two dimensional $SU(N)$ 
$(N \rightarrow \infty)$ gauge fields  interacting with 
matter in both adjoint and fundamental representations we investigate
the nature of the phase transition separating the strong and 
weak coupling regions of the phase diagram.  By interpreting the 
large $N$ solution of the model in terms of 
$SU(N)$ representations it is shown that the strong 
coupling phase corresponds to a region where a gap occurs
in the spectrum of irreducible representations.  
We identify a gauge invariant order parameter for the generalized 
confinement-deconfinement transition and
give a physical meaning to each phase in terms of the interaction
of a pair of test charges.

\end{abstract}
\end{center}
\newpage
\setcounter{page}1
\section{Introduction}

The study of systems where interactions are mediated by non-Abelian 
gauge fields is of direct relevance to the physically 
interesting case of quantum chromodynamics (QCD).  At high temperature
or density these systems are expected to undergo a phase
transition where the character of the effective degrees of 
freedom changes dramatically.  For example, in the low 
temperature phase of four dimensional QCD,
quarks and gluons carrying colour charge
are not observed but rather confined into composite baryons and mesons.  
It is expected, and can be shown in numerical
simulations on the lattice, that at sufficiently high temperatures
this confinement is relaxed and the fundamental degrees
of freedom become mobile in a quark-gluon plasma.
Quantifying the differences between these phases has been a subject
of study for some time now \cite{weiss} and is adequately
understood only in the 
case of pure gluo-dynamics without quarks.   Here the 
Polyakov loop operator \cite{polsus}
\beq
< \Tr ~g(x)> = < \Tr~P\e^{i \int_0^{1/T} A_0(x, \tau) d\tau}>
\label{ploop}
\eeq
provides an 
effective order parameter \cite{sy} for the transition from the confined 
to the deconfined phase by testing to see if the symmetries
of the action are realized faithfully. As we will show it is 
useful to consider the trace of the group element $g$ in 
group theoretic terms as defining a group character. Taking
$g$ in different irreducible representations will allow
us to unambiguously define the strong and weak coupling regimes
of a two dimensional model even in the presence of fundamental 
matter.

Solvable models are often of use in developing new ideas and 
testing hypotheses. Here we will use the model of the
non-Abelian Coulomb gas in one space and one time dimension which
has a non-trivial phase structure qualitatively similar to what we 
expect in four dimensional QCD.
In particular we will discuss a finite temperature 
model of static adjoint and 
fundamental representation charges interacting through
$SU(N)$ gauge fields in the limit $N \rightarrow \infty$ \cite{gps}.
It is most conveniently presented as a unitary matrix model and its
solution can be found explicitly in the large $N$ limit by
calculating the element $g_0$ of the gauge group which dominates a
saddle-point approximation of the partition function.
The phase structure of the model, as we will show, consists of two 
general regions, one which we will define as weak coupling where 
$g_0 \sim 1$ and one of strong coupling with $g_0$ away 
from the identity element. It was noted \cite{gps}
that the behaviour in the system of  
$\Tr ~ g_0^n$ is markedly different in each phase 
when varying the parameter $n$. Some effort was made to interpret these
characteristics physically and using the language of group characters
we complete that task here.

Even though two dimensional Yang-Mills is a dynamically trivial theory,
as the rank of the symmetry group is taken to infinity,
group theory can drive phase transitions.
Transitions of this type were first noted long ago in the 
lattice theory \cite{gw} but these are now considered to be lattice
artifacts.  More recently such phase transitions have been noted
in the continuum with the Douglas-Kazakov transition 
\cite{dougkaz}
on the sphere and the related transition on the cylinder
\cite{matgro} being prime examples.  In these cases the theory
is solved for large rank symmetry group in terms of 
a single  irreducible representation
which saturates the evaluation of the 
partition function in a saddle-point approximation.
The phase transition corresponds to a point where the distribution 
of occupation numbers for the rows of the associated
Young table develops a gap \cite{dougkaz,matgro}.  
 
In the present case under consideration the situation is somewhat
different. The saddle-point is not determined in general
by a single irreducible representation of the gauge group
but by a linear combination of irreducible 
representations. This feature is also shared by Abelian and non-Abelian
Coulomb gases in two dimensions with  $U(N)$ and $SU(N)$ finite 
rank gauge groups \cite{nambu}
and can be generalized to the case of any compact Lie gauge group.
In each of these cases the state vector of the system, $\Psi$
is a class function and therefore 
can be represented by a linear combination of characters, $\chi_R$
of the
irreducible representations, $R$ of the gauge group with coefficients 
$a_R$ that depend on the parameters of the model (temperature, pressure, 
gauge coupling constant...)
\beq
\Psi[g(x)] = \sum_{R} a_R \chi_R(g(x)) 
\eeq

Consequently we see that there are two different points of 
view to take with solving these models in two dimensions.
One is to find a dominant configuration of the gauge group $g_0$
and the other is to find the dominant linear combination of  irreducible 
representations, $\Psi$.  The main objective here
is to quantify the connection between these two views
and use it to characterize the 
differences between the strong and weak coupling 
regimes of the non-Abelian Coulomb gas.
As we will see, the 
characters of the gauge group are completely determined
by traces of powers of the gauge matrices, $\Tr ~g^n$.  
In the two dimensional model under consideration
we will show that the vanishing of particular 
coefficients $a_R$ provides a convenient way to characterize the 
different phases of the model. Clearly,
if a particular $a_R$ is vanishing
then the system does not have excitations which can effectively 
screen a charge in irreducible representation $R$ interacting 
with its conjugate $\bar{R}$.  In this way we will be able to 
identify an order parameter for the transition from strong to 
weak coupling and give a physical definition of the 
confinement-deconfinement transition with fundamental 
matter present.

The layout of this paper is as follows: First we present a 
short description of the non-Abelian Coulomb gas model which 
will be used as a test-bed for our program of using group characters
to describe the different phases of a system of interacting gauge 
fields. In particular we will show that this model possesses an interesting 
phase structure. We follow with an explicit demonstration of 
the connection between a gauge group element
$g$ and the characters of the irreducible
representations of the the gauge group.  Applying these
general results to the case of the model at hand we will show
that the spectrum of irreducible representations provides
a clear way to distinguish between 
phases of the model with fundamental matter in much the 
same way the Polyakov loop operator does in pure gluo-dynamics.

\section{Review of the Non-Abelian Coulomb Gas}
\setcounter{equation}{0}

The model on which we will base our investigation is that of the
non-Abelian Coulomb gas in two dimensions as previously studied
in \cite{gps}.  This model, while lacking both dynamical 
gauge  and matter degrees of freedom, exhibits many of the 
features expected \cite{banks}
in higher dimensional systems.  For instance
in $SU(N)$ gluo-dynamics for $N \ge 3$ we expect a first order 
deconfinement transition with or without adjoint matter.  As
one adds matter in the fundamental representation it is expected 
that the latent heat associated with this transition drops until 
the transition becomes continuous or is completely washed out.
As we will show in this section, the two dimensional  
non-Abelian Coulomb gas shares this behaviour and is explicitly
solvable in the large $N$ limit.
The fact that we consider the limit where the rank of the 
symmetry group is taken to infinity is necessary to generate a 
phase transition in this simple model.  

We will now outline the details of this model which are relevant
to the phase structure and our interpretation of it. 
Beginning with the canonical quantization of $1+1$ dimensional
Yang-Mills theory with 
gauge coupling $e$ at finite temperature $T$ it can 
be shown \cite{Zar,stz,gps} that  
the grand partition function for a system of interacting
static colour charges is that of a gauged principal chiral 
model
\begin{equation}
Z[\lambda,T]~=~\int [dA][dg]~
\exp\left[-\int dx\left(\frac{N}{2 \gamma}~ {\rm Tr}\left|
\nabla g + i[A,g]\right|^2 -\lambda\left| {\rm
Tr}~g\right|^2  - 2 N \kappa ~{\rm Re} \Tr ~g \right) \right]
\label{partition}
\end{equation}
Here the integration over  gauge fields $A$ effectively 
enforces Gauss' law as one integrates over all elements of the 
gauge group with the Haar measure $[dg]$.
The fugacities of the adjoint and fundamental charges are
given by the parameters $\lambda$ and $N \kappa$, respectively.
Since we consider the matrix-valued fields $A$ and $g$ to 
be taken in the fundamental representation of $SU(N)$, 
the large $N$ limit will lead directly to the familiar 
situation of matrix models with large $N\times N$ matrices.
In order to keep all terms in the action of (\ref{partition})
at leading, $N^2$, order in this limit we will 
restrict parameters of the system such that
$\gamma \equiv  \frac{2T}{e^2 N} $, $\lambda$ and $\kappa$ 
are each of $O(1)$.

For the discussion of a confinement-deconfinement phase transition 
the most important aspect of the action in (\ref{partition})
is a global symmetry $S[A,g]~=$ $ S[A,~z~g]$
when the fundamental charge fugacity $\kappa$ vanishes. 
Here $z$ is a constant
element from the center of the gauge group, which for U(N) is U(1) and
for SU(N) is $Z_N$. It is this symmetry and its (thermo-)dynamical 
breaking that leads to the deconfinement phase transition in this model.
If $\kappa \neq 0$ the question of what remnants of this symmetry
persist is one we will answer in the next sections.
  
Additionally, there is a gauge invariance that can be used to diagonalize 
the matrices $g_{ij}(x)= e^{i\alpha_i(x)}\delta_{ij}$. 
The density of eigenvalues
$\rho(\theta,x)~=~\frac{1}{N}\sum_{i=1}^N \delta(\theta-\alpha_i(x)) $
corresponding to the large $N$ saddle-point evaluation of 
(\ref{partition}) now completely characterizes the
properties of the system.  Our goal is to find  this 
distribution  of eigenvalues.
Without loss of generality we can consider the Fourier expansion
\begin{equation}
\rho(\theta,x)~=~\frac{1}{2\pi}+\frac{1}{2\pi}
\sum_{n\neq 0}c_n(x) e^{-in\theta}
~~~,~~c_n(x)^*= c_{-n}(x) 
\label{den}
\end{equation}
The configurations of the eigenvalue density
(\ref{den}) that saturate (\ref{partition}) at large $N$ can 
be found via the 
collective field theory approach \cite{JS,Zar}.   
The method is essentially based on the
relation between matrix quantum mechanics and non-relativistic
fermions \cite{bipz}.  Leaving the details to \cite{gps}, it can be
shown that a solution of the saddle-point evaluation
of (\ref{partition}) is given by
\begin{equation}
\rho_0(\theta)=\left\{ \matrix{
\sqrt{\frac{8}{\gamma\pi^2} }\sqrt{ E+ 2 (\lambda c_{1}+\kappa)
\cos{ \theta} } &
\mbox{where } \rho \mbox{ is real} \cr 0 & {\rm otherwise}\cr}\right.
\label{Cdens}
\end{equation}
The constant of integration $E$ has physical interpretation as the
Fermi energy of a collection of $N$ fermions \cite{bipz}
on the circle subject to a periodic potential. It
is fixed by requiring the eigenvalue distribution to have 
unit normalization.
Furthermore, since the potential due to adjoint charges is 
non-local in eigenvalue space, 
the Fourier moment $c_1$ (see (\ref{den})) must be
self-consistently determined \cite{gubkleb}
\begin{equation}
c_1 \; = \;  \int d \theta ~\rho_0(\theta) \cos{\theta} 
\label{cnphi}
\end{equation}
This pair of conditions is most conveniently analyzed by
introducing a new parameter $\mu = E/(2(\lambda c_{1}+\kappa))$ and
the integrals over the positive support of $\mu +\cos{\theta}$,
\beq 
I_n(\mu)=\frac{2}{\pi } \int d \theta \cos{n \theta} 
\sqrt{\mu +\cos{\theta}}
\eeq
In terms of $\mu$, the solution of the normalization and moment
conditions is given by
\beq 
c_1 = \frac{I_1(\mu)}{I_0(\mu)}
\eeq
and 
\begin{equation}
\frac{{\kappa}}{\gamma} \; \; = \; \;  
\frac{1}{4 I_0(\mu)^2} - \frac{\lambda}{\gamma} \frac{ I_1(\mu)}{I_0(\mu)} 
\label{Cnecc}
\end{equation}
This last relation gives a family of lines in the $(\lambda /\gamma, 
\kappa/ \gamma)$ plane parameterized by $\mu$. As is shown in 
Fig.1 this family overlaps itself for lower densities
of fundamental charges, $\kappa / \gamma$ signalling the fact that
there are multiple solutions to the equations of motion in this region
of the phase diagram.
Considering the free energy one can show \cite{gps} that for lower
densities of adjoint charges the stable solution has $\mu>1$ while
at higher adjoint densities the stable solution has $\mu<1$.
In the intermediate regime there is a first order phase transition.
As the density of fundamental charges is increased the first order 
transition is smoothed out and a third order phase transition 
persists along the line $\mu=1$.
\begin{figure}
\epsfysize=2in
\epsfbox [-150 245 530 535] {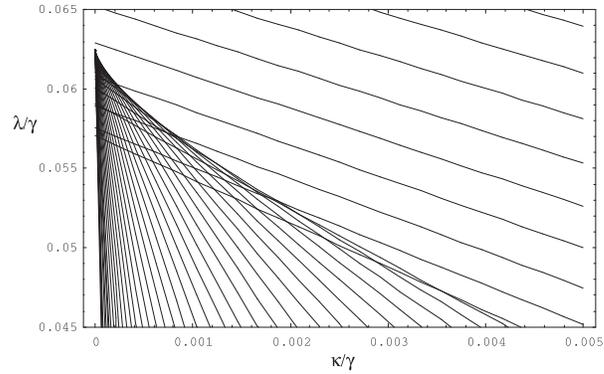}
\caption{{\it Plot of the lines (2.7) for $\mu$ ranging from 
$0.4$ (upper right corner) to $75$ (line at the extreme left).  
The region of overlapping lines corresponds to
a region of first order phase transition. \label{linesfig} }}
\end{figure}

The parameter $\mu$ is now seen to be useful for two different
reasons.  First it characterizes the general structure of the 
phase diagram (Fig. 2) where the `strong coupling' regime
is the region with $\mu>1$ and the `weak coupling' regime has
$\mu<1$.  As well, and of more importance for our analysis, 
we find that the expectation values of traces of powers of 
the group element $g$ are given as a function of the 
single parameter $\mu$
\beq
< \Tr~g^n /N> = c_n = \frac{I_n(\mu)}{I_0(\mu)}
\eeq
Consequently, it makes sense for our purposes to re-define 
the eigenvalue distribution in terms of $\mu$
\beq
\rho_0(\theta, \mu) = \frac{2}{\pi I_0(\mu)} 
\sqrt{ \mu + \cos{\theta}}
\label{rhomu}
\eeq
In the next section we will use this definition and its connection 
to the dominant configuration of the gauge element to 
analyze the phase diagram in terms of group theory.
\begin{figure}
\epsfxsize=4in
\epsfbox[-270 251 456 648]{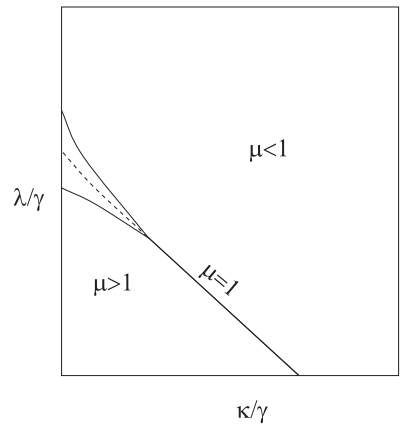}
\caption{{\it Schematic picture of the phase diagram. 
The doted curve marks the first order 
part of the critical line. The solid curves above and below it are the 
boundaries of the area with two possible phases. They join at a point
which shows second order behaviour. For larger $\kappa/\gamma$
we find a third order line ($\mu = 1$) marked by a solid line.
\label{phasediag}} }
\end{figure}

\section{Order parameters and some group theory}
\setcounter{equation}{0}

As is known, in the case of pure gluo-dynamics, 
the realization of the center symmetry of the gauge group
governs confinement ~\cite{polsus}.  The Polyakov loop
operator ${\rm Tr}~ g(x)$, which is related to the free energy
$-T \log{ <\rm{Tr} g(x) \rm{Tr} g^\dagger (0)>}$  of a 
conjugate pair of static, external fundamental charges  
separated by a distance $x$, serves as an order parameter 
\cite{sy} to test confinement. Since
${\rm Tr}~ g(x)$ transforms under the center as ${\rm
Tr}~g(x)~\rightarrow z~{\rm Tr}~g(x) $, the expectation value of
the Polyakov loop operator must average to zero if the 
center symmetry is preserved. Physically this suggests that
an infinite amount of energy is required to introduce
a single fundamental test charge into the system. The 
presence of a gas of fundamental charges   
$(\kappa \neq 0)$ changes this situation though by explicitly 
breaking the center symmetry. Consequently we lose the 
Polyakov loop operator as an order parameter for phase
transitions in the system. In this section 
we introduce a suitable
generalization of the Polyakov loop operator which will allow
us to identify a new order parameter.
 
As seen in the previous section, the
solution of the non-Abelian Coulomb gas with adjoint and 
fundamental representation charges is completely characterized 
by a Fourier sum of the 
traces $c_n= <\Tr~g^n/N>$ - the higher winding Polyakov loops.
As noted in \cite{gps} the character of 
these traces changes between the strong and weak coupling regimes.
In particular, in the strong coupling $(\mu >1)$
phase $c_n$ is damped exponentially with $n$ while in the 
weak coupling $(\mu <1)$
phase the damping follows a power law behaviour.
Now we will reconsider this behaviour in terms of 
group theory. 

Since the matrix $g$ is an element of the special unitary group, 
its trace in an irreducible representation, $R$  
defines the group character for that representation
\beq
\chi_R(g) \equiv \Tr_R~g
\eeq
For the $N$ dimensional fundamental representation of $SU(N)$, $F$,
the group
character is just the Polyakov loop operator described above
since we are considering group elements to be taken 
in the lowest fundamental representation
\beq
\chi_F(g) = \Tr ~g
\eeq
Further simple examples are
the symmetric $(S)$ and anti-symmetric $(A)$ combinations
of a pair of fundamentals where we have
\beq
\chi_S(g) =\frac{1}{2}[ (\Tr~ g)^2 + \Tr ~g^2 ]
~~,~~\chi_A(g) =\frac{1}{2}[ (\Tr ~g)^2 - \Tr ~g^2 ]
\eeq
A general relation between characters and the group elements
is given by the Weyl formula but is not necessary for the 
following. A complete discussion can be found in 
standard references (see \cite{grpth} for example).
 
The main idea is that the eigenvalues of the group matrices, which 
are the only relevant dynamical variables in the grand partition 
function (\ref{partition}), are completely determined by the
$N$ quantities $\{ \Tr ~ g^n \}$, $n =1 \ldots N $.  In turn these traces 
form an algebraic basis equivalent to the characters of the 
$N$ fundamental (completely anti-symmetric)
irreducible representations of $SU(N)$ (including the 
trivial representation). 
Here we will explicitly
demonstrate the relationship between the basis of traces and the 
basis of group characters.
Ultimately it is the group theoretic variables which we will
use to characterize the phases of the model (\ref{partition}).
 
The standard basis for general
functions (of finite degree) of the eigenvalues of a matrix
is the set of elementary symmetric functions $\{a_r \}$.
In terms  of the eigenvalues  $\lambda_j =\e^{i \theta_j}$
of the group element $g$ they are given by
\bea
a_1 &=& \sum_j \lambda_j \\
a_2 &=& \sum_{j<k} \lambda_j \lambda_k  \nonumber \\
a_3 &=& \sum_{j<k<l} \lambda_j \lambda_k \lambda_l  \nonumber \\ 
\vdots  \nonumber \\
a_N &=&  \prod \lambda_j =\det{g} = 1
\eea
with $a_r \equiv 0$ for $r > N$.
The relationship of the symmetric functions $\{ a_r \}$ to the 
traces of the group elements, $S_n =\Tr ~ g^n$, is given \cite{lwd}
by the determinant 
\beq
a_k = \frac{1}{k!} \left| \begin{array}{ccccccc}
S_1 & 1 & 0 &\cdots & & & \\
S_2 & S_1 & 2 & 0 &\cdots & & \\
S_3 & S_2 & S_1 & 3 & 0 &\cdots &  \\
\vdots & \vdots &\vdots &\vdots & & & 0 \\
S_{k-1}& S_{k-2} & S_{k-3} & \cdots & S_2& S_1& k-1 \\ 
S_k& S_{k-1}& S_{k-2} &\cdots & S_3& S_2& S_1   
\end{array} \right|
\label{bigdet}
\eeq
Most importantly, it
can be shown  that the elementary symmetric functions are
nothing more than the characters of the fundamental representations
for the unitary group \cite{lwd,hagen}.  That is, for the 
fundamental representation which is the anti-symmetric combination
of $k$, $N$ dimensional representations, $\chi_k (g) = a_k$.

The determinant (\ref{bigdet}) can be evaluated \cite{muir}
in terms of a multinomial expansion most
compactly stated in terms of a generating function 
\beq
\chi_{k}(g)=  \frac{(-1)^k}{k!} \frac{d^k}{dz^k}
\left. \exp{[ -\sum_{n=1}^{\infty} \frac{\Tr g^n}{n} z^n]}
\right|_{z=0}
\eeq
For our purposes though it is useful to convert to a contour integral
about the origin. 
\beq 
\chi_{k}(g) = \frac{(-1)^k}{2 \pi i } \oint \frac{dz}{z^{k+1}}
\exp{ \left[ -\sum_{n=1}^{\infty} \frac{\Tr g^n}{n} z^n \right]}
\label{genchar}
\eeq
These last two expressions explicitly 
demonstrate the relationship between the 
group element $g$ and the $k^{th}$ fundamental representation of the 
gauge group and are completely general results.

With these relations we see that there is a direct connection
between the gauge group element $g$ and the irreducible 
(fundamental) representations of the gauge group.
In particular, in the previous section we have seen that
in the large $N$ solution of the non-Abelian Coulomb gas
a certain configuration of the gauge matrix, $g_0$ saturated 
the evaluation of the partition function (\ref{partition}).
Now it is natural to ask what is the configuration of 
irreducible representations corresponding to the dominant $g_0$.
This corresponds to evaluating the expectation
$< \chi_k(g)>$ in the background of the non-Abelian gas.  
In principle this involves calculating expectations
of the form $< \Tr ~ g^{n_1} \cdots \Tr~ g^{n_r}>$ but because of the 
factorization of gauge invariant objects in the limit $N \rightarrow 
\infty$, this reduces to a product of expectations, 
$< \Tr ~ g^{n_1}> \cdots <\Tr~ g^{n_r}>$.  Consequently 
$< \chi_k(g)>$ is determined by replacing $\Tr~ g^n$
by its expectation value in (\ref{genchar}).  
Of course expectation values of the group element traces  are
intimately related to the eigenvalue density $\rho(\theta,\mu)$
(see (\ref{den})) hence, after performing an infinite sum, we obtain
\beq
<\chi_{\alpha}>[ \rho(\theta, \mu) ] \equiv
\frac{(-1)^{\alpha N}}{2 \pi i } \oint \frac{dz}{z}
\exp{ \left[ \frac{N}{2} \int d \theta \rho(\theta,\mu) 
\log{\left(\frac{ 1 + z^2 - 2 z \cos{\theta}}{z^{2 \alpha}} \right)}
\right]}
\label{charint}
\eeq
Note that we have defined a new real parameter $\alpha= k/N$ on the 
unit interval that effectively 
labels the fundamental representations in the large $N$ limit.
Of course (\ref{charint}) now depends on a continuous variable
and is of a slightly different functional form than the discrete
case $<\chi_k>$.  In the remainder of this discussion we will 
consider only the character parameterized by $\alpha$ as 
defined in (\ref{charint}).

\section{Calculation of the expectation of fundamentals}
\setcounter{equation}{0}
 
In this section we will concentrate on calculating 
$<\chi_{\alpha}>$ 
with eigenvalue density (\ref{rhomu}) for the non-Abelian Coulomb gas.
This calculation will give a clear picture of the 
group theoretic excitations present in different regions of 
the phase diagram and consequently allow us to define 
an order parameter for the deconfinement transition even 
in the presence of fundamental matter.

Since explicit evaluation of (\ref{charint}) is difficult 
we begin with some special limiting cases.
As $\mu \rightarrow -1$ the support of the 
eigenvalue distribution (\ref{rhomu})
vanishes at $\theta =0$. The distribution does not vanish 
though as it retains
unit normalization and effectively becomes a delta function, 
$\delta(\theta)$.
Consequently we find the gauge matrix $g$ is just the identity
at $\mu=-1$, hence
\beq
< \chi_\alpha > = \lim_{N \rightarrow \infty}
\left( \begin{array}{c} N \\ \alpha N
\end{array} \right) =
2^N \sqrt{ \frac{2}{N \pi}} 
\e^{-2 N (\alpha -1/2)^2}
\eeq
In this limit we find that the distribution of characters is 
symmetric about $\alpha =1/2$ as one would expect in a system 
where the total colour charge is vanishing.  As well in this limit
$< \chi_\alpha >$ is non-vanishing and 
all fundamental representations are present in 
the large $N$ background solution of the model.  
As we will see, this result
is generic in the weak coupling phase $\mu <1$.

In the opposite limit, as $\mu \rightarrow
\infty$, it can be shown that the eigenvalue distribution 
(\ref{rhomu}) approaches a constant value $\rho=1/2 \pi$ with 
the eigenvalues of the group element $g$
becoming uniformly distributed on the unit circle.
Since expectation values of the traces of powers of 
the gauge matrix are essentially Fourier transforms of 
the eigenvalue distribution, it is easy to see that
$<\Tr~g^n> \rightarrow 0$ in this limit and 
\beq
< \chi_\alpha > \rightarrow \delta_{0,\alpha }
\eeq
This limit corresponds to the extreme strong coupling phase
of the model where the Polyakov loop operator $(<\Tr~g> \sim 
< \chi_{1/N} > )$ has vanishing 
expectation value and the standard analysis would point to 
a phase where colour charges are strictly confined into 
hadron-like structures.

In general the integral (\ref{charint}) can be evaluated by saddle-point
methods in the large $N$ limit in which we are interested.
The relevant action in this limit is
\beq
S( \alpha, \mu, z)=\int d \theta \rho(\theta, \mu) 
\log{\left(\frac{ 1 + z^2 - 2 z \cos{\theta}}{z^{2 \alpha}} \right)}
\label{charS}
\eeq
Solving the stationarity condition, $dS/dz |_{z_0} =0$,    
for $\alpha$ in terms of $z_0$ we find
the saddle-point condition for the large $N$ behaviour of the 
integral (\ref{charint}) is given by the relationship 
\beq
\alpha = \int d \theta \rho(\theta, \mu) \frac{z _0 (z _ 0 - \cos{\theta})}
{1 + z_0^2 - 2 z_0 \cos{\theta}} 
\label{saddlept}
\eeq
Since $\alpha$ is a real parameter restricted to 
the unit interval $[0,1]$ it can be shown 
that the saddle-point value of the parameter $z_0$ is real.
Further, for $z_0>1$  and $0<z_0<1$ Eqn.\ref{saddlept} returns values
of  $\alpha >1$ and  $\alpha <0$, respectively.  
Consequently we need only consider real, negative values of the 
parameter $z_0$. 

We now turn to an examination of the saddle-point approximation of
(\ref{charint})
for different regions of the phase diagram of the model at hand
beginning with the weak coupling phase, $\mu<1$. 
In this case the support of the 
eigenvalue distribution (\ref{rhomu}) is bounded away from 
$\theta =\pm \pi$ and hence the denominator in (\ref{saddlept})
is non-singular for all values of $z_0$. Consequently, in this regime
$\alpha$ varies smoothly and monotonically with $z_0$ and the relation (\ref{saddlept}) can in principle be inverted to obtain $z_0(\alpha)$.
With this information, the large $N$ 
asymptotic form of the expectation value of the 
characters $< \chi_{\alpha} >$ can be determined by standard saddle-point 
methods.  In Figure 3 we show a numerically calculated example
of $\alpha$ as a function of $z_0$ for 
$\mu=0.5$. For this same case we
show a schematic diagram of the magnitude of the expectation value
$|< \chi_{\alpha} >|$ as a function of $\alpha$ in Figure 4.
In particular we see that the system has excitations in all irreducible 
representations.

\begin{figure}
\epsfysize=2in
\epsfbox[-100 327 398 535]{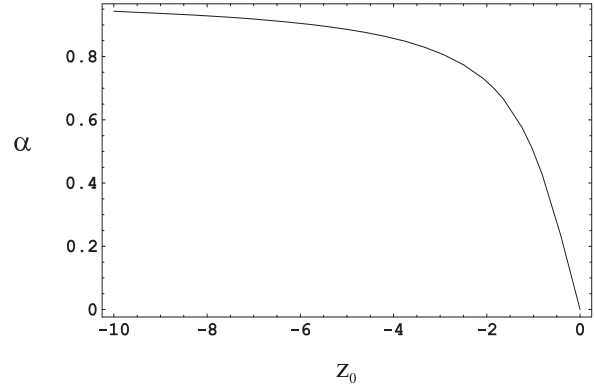}
\caption{ {\it Saddle-point relation for $\mu=0.5$ }}
\end{figure}

\begin{figure}
\epsfysize=2in
\epsfbox[-10 339 462 533]{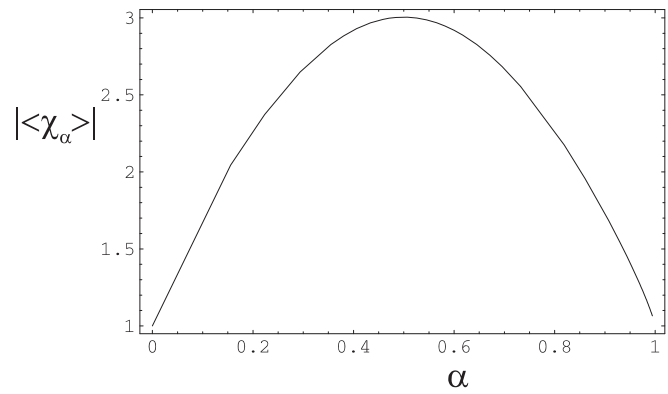}
\caption{{\it 
Schematic diagram of $|<\chi_\alpha >|$ vs. $\alpha$ for $\mu=0.5$ }}
\end{figure}

For $\mu>1$ the situation is somewhat different. Now the 
support of the eigenvalue distribution (\ref{rhomu}) is 
the full interval $\theta \in [-\pi,\pi]$, and the denominator of 
(\ref{saddlept}) causes non-analytic behaviour to appear.
As one increases 
$\mu$ through unity the saddle-point relation for $\alpha$ shows this
non-analytic behaviour as 
a discontinuity at $z_0=-1$ (see Figure 5). 
The result is that  
an open interval of $\alpha$ values centered on 
$\alpha =1/2$ are mapped into this discontinuity when the 
saddle-point relation (\ref{saddlept}) is inverted.  Since this 
discontinuity occurs in the saddle-point relation, it is 
not surprising to find that the curvature associated with 
the Gaussian integration of the saddle-point approximation
is divergent, effectively forcing the integral 
to vanish.  In terms of the expectation values of different
representations in the background of the non-Abelian Coulomb gas,
we see that an open interval of fundamental
representations centered about $\alpha =1/2$ is missing from 
the spectrum in the large $N$ limit.
In Figure 6 we show an example 
of the behaviour of the expectation value
$|< \chi_{\alpha} >|$ with $\alpha$ for $\mu=1.2$.
 
\begin{figure}
\epsfysize=2in
\epsfbox[-40 354 429 551]{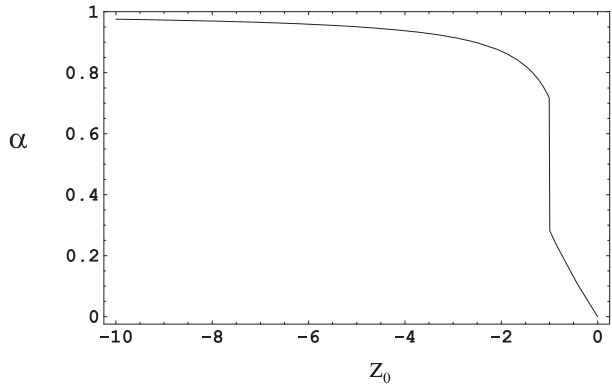}
\caption{{\it Saddle-point relation for $\mu=1.2$ }}
\end{figure}

\begin{figure}
\epsfysize=2.2in
\epsfbox[-40 329 451 551]{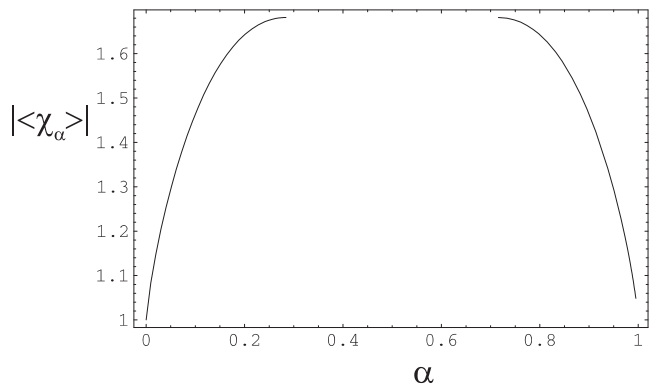}
\caption{{\it 
Schematic diagram of $|<\chi_\alpha >|$ vs. $\alpha$ for 
$\mu=1.2$ }}
\end{figure}

The main outcome of this analysis is that the expectation value
of the central fundamental character $<\chi_{1/2}>$ is vanishing 
if and only if $\mu \ge 1$. Consequently it may be considered
an order parameter distinguishing
between the strong and weak coupling phases of the model.  
Physically the situation is clear.  In the weak coupling phase
the system can effectively screen the interactions of any pair of
charges regardless of their representation since the system contains
excitations in all representations of the gauge group.
We conclude that the system 
looks much like a quark-gluon plasma where charges are effectively 
deconfined.  At the phase transition line
non-Abelian flux in the $\alpha=1/2$ fundamental representation 
becomes too energetically costly to produce and
the system can no longer screen the interaction between a pair
of $\alpha=1/2$ fundamental charges. In this strong coupling phase 
the interacting pair sees a linear confining potential (though
somewhat reduced as compared to the empty background).
As one further increases $\mu$ the
gap in the spectrum of fundamental representations becomes larger
and in the extreme limit $\mu = \infty$ the system contains only
excitations in the trivial representation.  This is precisely the 
confining phase of pure gluo-dynamics.

\section{Discussion}

As we have shown, the generalization of the concept of the 
Polyakov loop operator to probe the group theoretic 
excitations of a system of non-Abelian electric charges
provides a convenient and unified way to quantify the
physics of phase transitions.
While the details of our presentation
have centered on a two dimensional model with an infinite
number of colours, the general concepts developed here should be 
applicable to interacting gauge systems in arbitrary
dimensions for both infinite and finite rank $(N)$ gauge groups.
Unfortunately we are 
unable to test these ideas in a solvable two dimensional model
as for finite $N$ there is no phase transition and 
the system is always in the deconfined phase where all 
representations are present.

One immediate problem with 
using the fundamental representations to characterize the phase
diagram  arises when considering finite, odd rank groups.  
For example in the physically 
relevant case of $SU(3)$, 
there are only two fundamental representations $k=1,2$ and the 
order parameter $< \chi_{1/2}>$ 
would naively denote the $k=3/2$ fundamental 
representation.  In terms of group theory this fractional
representation is nonsense 
and strongly suggests that direct application of the large $N$
results is not prudent.  Even worse is the fact that a finite number
of representations are not enough degrees of freedom to describe
the change in character of the group elements from $g \sim 1$ in 
the weak coupling phase to $g$ away from the identity in the 
strong coupling phase.
Alternatively, we are free to use 
any independent set of irreducible representations
to characterize the system. In particular
the completely symmetric representations provide an equivalent 
algebraic basis to the fundamentals we have considered here.
The strength of this approach is that there is no restriction 
on the number of symmetric representations for the unitary groups
contrasting the $N-1$ fundamental representations for $SU(N)$.
Unfortunately,
repeating the calculations of Section 4 with symmetric
representations one discovers that there is no strong signal
for the phase transition and in particular one cannot define
an order parameter.

Despite the apparent difficulties with applying the current
results to finite rank gauge theories, we feel that the 
general concept
of characterizing the phases of a gauge theory coupled to matter
by the spectrum of higher irreducible representations is useful.
In particular it would be interesting to investigate
these ideas in the setting of lattice calculations of 
gauge theories that are known to possess phase transitions.
In fact, for the case of $SU(2)$ gauge theory, the 
$J=1/2,1$ and $3/2$ Polyakov loops have been calculated on the lattice
\cite{kiskis} (see also \cite{dh}). It
would certainly be very instructive to have more complete information
about the higher representation Polyakov loops in this  
simple model both with and without fundamental representation matter.

\section*{Acknowledgements}
We would like to thank G. Semenoff, A. Zhitnitsky, C. Gattringer 
and I. Halperin for helpful discussions.

\end{document}